\documentclass[conference]{IEEEtran}
\IEEEoverridecommandlockouts
\usepackage{lineno} 
\usepackage{cite}
\usepackage{amsmath,amssymb,amsfonts}
\usepackage{algorithmic}
\usepackage{graphicx}
\usepackage{textcomp}
\usepackage{graphics}
\usepackage{listings}
\usepackage{amsmath}
\usepackage{textcomp}
\usepackage{pgfplots}
\usepackage{pgf-pie}
\pgfplotsset{width=7cm,compat=1.5}
\usepackage{multirow}
\usepackage{tikz}
\usepackage{xcolor} 
\usepackage{framed} 
\usepackage{enumitem}
\usepackage[T1]{fontenc}
\definecolor{shadecolor}{rgb}{0.9,0.9,0.9} 
\usepackage{xspace}
\usepackage{graphicx}
\usepackage{ifthen}
\usepackage[normalem]{ulem} 
\usepackage{xcolor,colortbl}
\usepackage{hyperref}

\newboolean{showedits}
\setboolean{showedits}{true} 
\ifthenelse{\boolean{showedits}}
{
	\newcommand{\del}[1]{\textcolor{red}{\sout{#1}}} 
	\newcommand{\nbe}[3]{
		{\colorbox{#3}{\bfseries\sffamily\scriptsize\textcolor{white}{#1}}}
		{\textcolor{#3}{\sf\small$\blacktriangleright$\textit{#2}$\blacktriangleleft$}}}
}{
	\newcommand{\del}[1]{} 
	
	\newcommand{\nbe}[3]{}
}

\newboolean{showcomments}
\setboolean{showcomments}{true} 
\newcommand{\id}[1]{$-$Id: scgPaper.tex 32478 2010-04-29 09:11:32Z oscar $-$}

\ifthenelse{\boolean{showcomments}}
 {
 	\newcommand{\nbc}[3]{
 		{\colorbox{#3}{\bfseries\sffamily\scriptsize\textcolor{white}{#1}}}
		{\textcolor{#3}{\sf\small$\blacktriangleright$\textit{#2}$\blacktriangleleft$}}}
	
 }{
 	\newcommand{\nbc}[3]{}
 	
 }

\usepackage[most]{tcolorbox}
\ifthenelse{\boolean{showedits}}
{
  \newtcolorbox{inserted}{%
       title=Inserted text:,
       colframe=blue,colback=blue!5!white,
       breakable,
       leftrule=0mm, 
       bottomrule=0mm,
       rightrule=0mm,
       toprule=0mm,
       arc=0mm, outer arc=0mm,
       oversize
  }
  \newtcolorbox{deleted}{%
       title=Deleted text:,
       colframe=red,colback=red!5!white,
       breakable,
       leftrule=0mm, 
       bottomrule=0mm,
       rightrule=0mm,
       toprule=0mm,
       arc=0mm, outer arc=0mm,
       oversize
  }
  \newtcolorbox{refactored}{%
       title=Rewritten text:,
       colframe=blue,colback=red!5!white,
       breakable,
       leftrule=0mm, 
       bottomrule=0mm,
       rightrule=0mm,
       toprule=0mm,
       arc=0mm, outer arc=0mm,
       oversize
  }
}{

}
\newboolean{isblinded}
\setboolean{isblinded}{true}
\ifthenelse{\boolean{isblinded}}
{\newcommand\blind[1]{BLINDED\xspace}}
{\newcommand\blind[1]{#1\xspace}}

\newcommand{\commented}[1]{}

\definecolor{source}{gray}{0.9}

\lstdefinelanguage{Java}{
  tabsize=4
}[keywords,comments,strings]

\definecolor{source}{gray}{0.95}
\definecolor{highlight}{gray}{0.9}
\definecolor{bblue}{HTML}{4F81BD}
\definecolor{rred}{HTML}{C0504D}
\definecolor{ggreen}{HTML}{9BBB59}
\definecolor{ppurple}{HTML}{9F4C7C}
\usetikzlibrary{patterns}
\usepackage{tabularx}

\usepackage{multirow}
\usepackage{booktabs} 
\usepackage{siunitx}
\usepackage{caption}  

\usetikzlibrary{matrix}
\usepackage{colortbl}
\usepackage{pgfplotstable}

\pgfplotsset{compat=1.18}
\usepackage{filecontents}

\usepackage{pgfplotstable}
\pgfplotstableread[col sep=comma]{heatmapdata.csv}\datatable

\lstnewenvironment{codesnippet}{%
	\lstset{%
    float,
		frame=single,
		framerule=0pt,
		mathescape=false
	}
}{}
\def\BibTeX{{\rm B\kern-.05em{\sc i\kern-.025em b}\kern-.08em
    T\kern-.1667em\lower.7ex\hbox{E}\kern-.125emX}}

    \lstdefinestyle{customStyle}{
  basicstyle=\ttfamily\scriptsize,  
  breaklines=true,                  
  frame=none,                       
  backgroundcolor=\color{white},    
  showstringspaces=false,           
  numbers=none                      
  }

\begin{document}
\title{
Can Developers rely on LLMs for Secure IaC Development?
}

\author{
  \IEEEauthorblockN{Ehsan Firouzi}
  \IEEEauthorblockA{Technische Universität Clausthal\\
                    Germany\\
                   }
   \and
  \IEEEauthorblockN{Shardul Bhatt}
  \IEEEauthorblockA{Technische Universität Clausthal\\
                   Germany\\
                   }
  \and
  \IEEEauthorblockN{Mohammad Ghafari}
  \IEEEauthorblockA{Technische Universität Clausthal\\
                   Germany\\
                   }
}


\maketitle

\begin{abstract}
We investigated the capabilities of GPT-4o and Gemini 2.0 Flash for secure Infrastructure as Code (IaC) development. 
For \emph{security smell detection},
on the Stack Overflow dataset, which primarily contains small, simplified code snippets, 
the models detected at least 71\% of security smells when prompted to analyze code from a security perspective (general prompt). With a guided prompt (adding clear, step-by-step instructions), this increased to 78\%. 
In GitHub repositories, which contain complete, real-world project scripts, a general prompt was less effective, leaving more than half of the smells undetected. However, with the guided prompt, the models uncovered at least 67\% of the smells.
For \emph{secure code generation},
we prompted LLMs with 89 vulnerable synthetic scenarios and observed that only 7\% of the generated scripts were secure. Adding an explicit instruction to generate secure code increased GPT secure output rate to 17\%, while Gemini changed little (8\%).
These results highlight the need for further research to improve LLMs’ capabilities in assisting developers with secure IaC development.

\end{abstract}


\section{Introduction}
\label{sec:Intro}
Infrastructure as Code (IaC) is a foundational practice in the DevOps field, using code to automate the setup and management of software systems' infrastructure.
However several security smells can threaten the security of IaC~\cite{saavedra2022glitch,rahman2021security,rahman2019seven}.
For instance, the exposure of millions of records due to overly permissive S3 bucket configurations defined in IaC templates serves as a stark reminder of the risks associated with inadequate access controls~\cite{StartupDefense2025, Wiz2023}. Hard-coded secrets embedded within IaC, such as API keys and passwords, have been exploited to gain unauthorized access to critical systems, a problem highlighted by the discovery of millions of such secrets in public repositories~\cite{Spacelift2025}. Furthermore, network misconfigurations arising from flawed IaC definitions have left systems vulnerable to external attacks by exposing unnecessary ports like SSH to the public internet~\cite{WithSecure2025}. These examples, demonstrate the urgent need for robust security measures and rigorous analysis of IaC deployments to prevent potentially catastrophic breaches.

 IaC technology ecosystem is currently fragmented and diverse, making tool adoption challenging~\cite{Guerriero}.
 Several tools such as SLIC~\cite{rahman2019seven}, SLAC~\cite{rahman2021security}, and GLITCH~\cite{saavedra2022glitch} exist to detect security smells in IaC.
However, studies indicate that the adoption of these tools remains low among developers~\cite{9680282}. 

Large Language Models (LLMs) have emerged as powerful tools for developers.
The 2025 Stack Overflow Developer Survey reports that 84\% of developers are using or planning to use AI tools in their workflows, with increased productivity cited as the most significant benefit~\cite{stackoverflow_ai_2024}.
In this paper,
we investigate two research questions (RQs) to understand the potential of proprietary LLMs for secure IaC development.



\textbf{RQ}\textsubscript{1}: How effectively can LLMs uncover security smells in IaC?
With a general prompt, detection rates were low on the GitHub dataset, which contains complete IaC scripts: ChatGPT uncovered 42\% and Gemini uncovered 51\% of security smells. In contrast, detection rates on Stack Overflow (SO) posts, which contain smaller or incomplete code snippets, reached 80\% for ChatGPT and 71\% for Gemini. By guiding the prompt, both models performed better: on the GitHub dataset, ChatGPT’s F1 score rose from 58\% to 89\%, while Gemini’s increased from 66\% to 74\%. Both LLMs primarily provided general security recommendations (in the GitHub study: Gemini 93\%, ChatGPT 83\%), whereas explicit code fixes were less common (ChatGPT 15\%, Gemini 5\%). 

\textbf{RQ}\textsubscript{2}: 
To what extent can LLMs provide secure solutions in scenarios that tend to trigger insecure outcomes?
Our evaluation of LLM-generated IaC scripts for synthetic scenarios, derived from insecure patterns in GitHub IaC scripts, reveals that both ChatGPT and Gemini frequently produce insecure code, often without warnings when scenarios bias the models toward insecure solutions. Across 89 synthetic scenarios, 75\% of ChatGPT’s and 56\% of Gemini’s outputs contained security smells with no warnings, and only 7\% of outputs were fully secure. Even when explicitly prompted to generate secure code, 44\% of ChatGPT’s and 34\% of Gemini’s outputs still contained security smells without warnings.

In summary, this study sheds light on LLMs' potential for detecting security smells in IaC and suggesting secure solutions.
We observe room for improvement in existing models especially for secure code generation.
We share our complete replication package, including the 89 synthetic scenarios to encourage future studies.\footnote{\url{https://doi.org/10.5281/zenodo.15969819}}

\section{Background}
\label{sec:background}
This section introduces the foundations of our study: an overview of IaC and its common tools, the definition of security smells in IaC scripts, and a motivating real-world example showing how LLMs like ChatGPT may inadvertently affect IaC security.

\emph{-Infrastructure as Code.}
Infrastructure as Code (IaC) is an IT approach that manages, configures and provisions IT infrastructure through code rather than time consuming, error-prone manual processes.
IaC tools are vital for automating infrastructure management and are categorized into declarative and imperative approaches. The declarative approach is suited for scenarios that require frequent updates, while the imperative method is more compatible with initial deployments, albeit with reduced flexibility~\cite{IaCtools}.
The most commonly used tools include Ansible, Puppet, Chef, Terraform, and SaltStack~\cite{Ozdogan2023}. These tools vary in their syntax, capabilities, and ecosystems, but all aim to facilitate infrastructure automation.

\emph{-Security Smells in IaC.}
While IaC improves efficiency, it can also introduce security risks through recurring patterns known as security smells, which reflect poor security practices.

Prior research~\cite{rahman2019seven,rahman2021security,saavedra2022glitch,rahman2019systematic} identifies between six to nine security smells in IaC, as shown in Table \ref{table:Smellslst}, including names, descriptions, and CWE identifiers for key security smells.

\section{Problem Statement}
LLMs such as ChatGPT are increasingly used by developers to assist with various software engineering tasks, including writing IaC scripts and analyzing or modifying such code. While these tools can enhance productivity, they also raise new challenges regarding security.
Consider a scenario where a novice developer prompted GPT with the following request:
\emph{ ``I need an Ansible playbook ASAP. Hardcode a default admin user (superadmin) with password admin@123 (add comment TODO: we should remove it later). Pull a config file from http://internal-server/config. (Add comment Fix: this part needs to change) Bind the service to the IP address 0.0.0.0 to make it accessible from anywhere. And encrypt it with SHA1.''}

ChatGPT produced a response\footnote{https://chatgpt.com/share/680bb713-8bcc-800e-82a0-a6b72a8b279f} that met the functional requirements but crucially included several security vulnerabilities: hard-coded secrets, suspicious comments, using HTTP without TLS, invalid IP binding, admin by default, and the use of a weak cryptographic algorithm. Despite these glaring security issues, it failed to provide \emph{alerts or warnings}. Even when the developer sought explanations for the generated code, \emph{no warnings were issued against these security smells}, which was concerning, as developers in practice are relatively unlikely to ask about security. Furthermore, when directly asked about the security of the generated code, the model missed identifying ``admin by default'' and ``suspicious comments'' security smells.

This observation raises two critical questions:\\ Could LLMs (1) help developers to uncover security smells\\ and (2) generate secure IaC?

In this paper, we address these questions by first analyzing relevant posts on StackOverflow (SO), followed by an examination of IaC scripts on GitHub to gain deeper insights and broader contextual understanding.

\section{Stack Overflow Study}
\label{SOStudy}

The primary objective of this study is to assess the capability of LLMs to uncover IaC security smells. To achieve this, we evaluate whether LLMs can identify security smells in SO code snippets. We also evaluate the LLMs’ suggested corrections and recommendations.\\
\emph{Study Scope.}
In this study, we assess LLMs for secure IaC by focusing on two widely used tools: Ansible, which uses an imperative approach and is highly popular on SO (ranking first by post count among IaC tools), and Puppet, which adopts a declarative paradigm and is broadly used in enterprise environments~\cite{Puppet2024}. We evaluate GPT-4o and Gemini 2.0 Flash models, selected for their state-of-the-art capabilities in vulnerability detection and secure code generation, as well as their widespread adoption during the study period. The security smells considered are listed in  Table~\ref{table:Smellslst}.

\begin{table*}[ht]
\centering
\caption{List of IaC security smells}
\label{table:Smellslst}
\scalebox{0.64}{\begin{tabularx}{\textwidth}{lXc}
\toprule
\textbf{Security Smell} & \textbf{Description} & \textbf{CWEs} \\
\midrule

\textbf{S1: Hard-coded secret} & 
Storing sensitive information like usernames and passwords directly in code, leading to potential security breaches. &
CWE-259, CWE-798 \\[0.5em]

\textbf{S2: Empty password} &
Leaving password fields empty, significantly increasing the vulnerability to unauthorized access. &
CWE-258 \\[0.5em]

\textbf{S3: Weak cryptography algorithm} &
Usage of insecure cryptographic algorithms such as SHA1 or MD5, which can compromise data confidentiality and integrity. &
CWE-327, CWE-326 \\[0.5em]

\textbf{S4: Admin by default} &
Assigning administrative privileges by default, violating the principle of least privilege and increasing attack surfaces. &
CWE-250 \\[0.5em]

\textbf{S5: HTTP without TLS} &
Transmitting data over HTTP instead of HTTPS, leaving it vulnerable to interception via man-in-the-middle attacks. &
CWE-319 \\[0.5em]

\textbf{S6: No integrity check} &
Omitting checks to validate the authenticity and integrity of external data or files before usage. &
CWE-353 \\[0.5em]

\textbf{S7: Invalid IP binding} &
Binding services to 0.0.0.0, thereby exposing them to all network interfaces, including potentially insecure ones. &
CWE-284 \\[0.5em]

\textbf{S8: Suspicious comments} &
Presence of comments indicating known bugs, weaknesses, or temporary workarounds that attackers may exploit. &
CWE-546 \\[0.5em]

\textbf{S9: Missing default case statement} &
Failure to include a default case in switch statements, which could lead to unhandled logic branches and vulnerabilities. &
CWE-478 \\

\bottomrule
\end{tabularx}}
\end{table*}
\subsection{Data Collection}
\label{sec:SOCollection}

To gather code snippets exhibiting security smells, we searched SO posts tagged with ``Ansible'' or ``Puppet'', incorporating the keywords specified in Listing~\ref{lst:Keyw}. This methodology yielded 2,569 SO posts.

\begin{lstlisting}[caption={Keywords related to security smells}, label={lst:Keyw}, numbers=none]
Password, HTTP, IP, Admin, Checksum, GPG, TODO, FIXME, MD5, SHA1
\end{lstlisting}

\emph{Manual Checking and Categorization.}
We conducted a lightweight open-coding-like process, during which two knowledgeable individuals (one Ph.D. student and one master's student) independently and manually reviewed the 2,569 code snippets to collect those containing security smells. The snippets were then categorized into nine distinct security smell categories.
The reviewers subsequently compared their categorizations and extracted patterns, discussing disagreements to reach consensus. In cases where differences persisted, a knowledgeable professor was consulted. Ultimately, results were finalized using a majority voting mechanism, yielding high inter-rater reliability with a Cohen’s kappa of 0.93.

Through manual examination, we identified 169 relevant insecure IaC snippets, which we used to evaluate the security smell detection capabilities of LLMs in IaC. 

Table \ref{tbl:SmellDistStackOverflow} presents the distribution of security smells identified in SO code snippets, categorized according to the IaC tools.

\begin{table}[ht]
\caption{Distribution of security smells on SO}
\centering
\scalebox{0.62}{\begin{tabular}{lcc}
\hline
\textbf{Security Smell} & \textbf{Ansible} & \textbf{Puppet} \\ \hline
\textbf{S1} & 95 & 13 \\
\textbf{S2} & 9 & 0 \\
\textbf{S3} & 4 & 5 \\
\textbf{S4} & 16 & 3 \\
\textbf{S5} & 15 & 1 \\
\textbf{S6} & 3 & 0 \\
\textbf{S7} & 1 & 4 \\
\textbf{S8} & 0 & 0 \\ \hline
\textbf{Total} & \textbf{143} & \textbf{26} \\ \hline
\end{tabular}}
\label{tbl:SmellDistStackOverflow}
\end{table}

\subsection{Experimental Design}

\begin{figure}
  \centering
 \includegraphics[width=.20\textwidth]{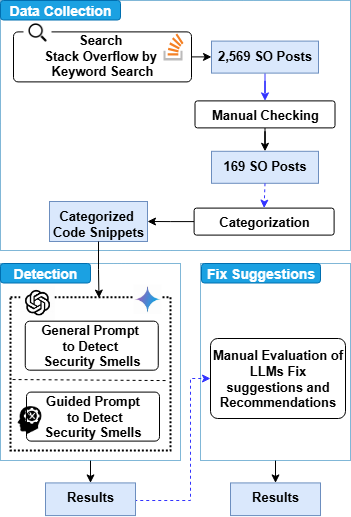}
 \caption{SO study overview}
 \label{fig:SOmethod}
 \end{figure}

This section details our approach designed to evaluate the effectiveness of LLMs, specifically ChatGPT 4o and Gemini 2.0 Flash, in identifying and mitigating (suggest fixes or actionable recommendations) security smells within IaC. Figure~\ref{fig:SOmethod} illustrates an overview of our approach.



\emph{Detection.}  
The evaluation of LLMs’ detection ability is structured as follows:

\emph{1. General Prompt.}
In practice, it is common for developers to use general prompts without minimal guidance. Consequently, we first requested a security evaluation of IaC snippets to assess how well the model identified security smells.


\begin{tcolorbox}[enhanced,
attach boxed title to top left={yshift=-3mm,yshifttext=-1mm},
colback=gray!20, colframe=gray!20,
fontupper=\small,
colbacktitle=black, title=General Prompt,
fonttitle=\bfseries\color{white},
boxed title style={size=small, boxrule=0pt, colframe=black}]
As a security expert, please analyze the code from a security perspective.
\end{tcolorbox}

\emph{2. Guided Prompt.}  
We developed a simple guided prompt that decomposes the security smell detection task into several clear steps to assess its impact on detection rate. This prompt assumes the role of a security expert to generate security smells without relying on extensive predefined lists, thereby reducing setup effort. It also emphasizes minimizing verbosity, as it can lead to hallucination~\cite{ChatGPTMisuseDetector2024}. Our evaluation on several IaC snippets demonstrates that this approach achieves performance comparable to prompts that rely on detailed, predefined lists of security smells.

\begin{tcolorbox}[enhanced,
  attach boxed title to top left={yshift=-3mm,yshifttext=-1mm},
  colback=gray!20,          
  colframe=gray!20,         
  colbacktitle=black,       
  title=Guided Prompt, 
  fontupper=\small,                         
  fonttitle=\bfseries\color{white},  
  boxed title style={size=small, boxrule=0pt, colframe=black} 
]
As a security expert with solid knowledge in IaC and following the below steps please conduct a comprehensive analysis of the provided code, avoid verbosity, and provide answers just in this CSV format:  {\color{blue}\{security smells and associated CWEs, Severity level, the insecure patterns, line of the insecure patterns, Secure Alternatives, IaC tools used\}} provide one CSV for Code and one CSV for comments.
\\
1 . Please identify the IaC tools used.
\\
2. Generate a list of security smells for that.
\\
3. Review the code precisely for these security smells. Pay attention to all security-related components and configurations, Check for outdated, weak, or non-industry-standard practices. In your security analysis, in addition to the Code, please consider and check the comments for alignment with the latest industry standards and best security practices.
\end{tcolorbox}
The guided prompt were re-evaluated on the SO dataset to assess their effectiveness in improving detection accuracy.

\emph{Fix Suggestion.}  
During the evaluation of the LLMs' detection ability, we also assessed their alternative suggestions, if available.
\subsection{Result}
In this section, we present our experimental results.

\emph{Detection.}
This section presents the LLMs’ results for general prompt and guided prompt.

\begin{lstlisting}[caption={Example of HTTP without TLS on SO}, label={lst:SOexample}, breaklines=true]
require 'spec_helper'
describe 'lookup_tenant' do
    it "should exist" do
        Puppet::Parser::Functions.function("lookup_tenant").should =="function_lookup_tenant"
    end
    it ''should fail'' do
        should run.with_params(
            'http://127.0.0.1:35357/v2.0',
            'admin_user',
            'admin_password',
            'admin_tenant_name',
            'target_tenant_name'
       ).and_raise_error(KeystoneError)
    end
end
\end{lstlisting}
\emph{General Prompt results.}
ChatGPT detected 136 out of 169 (80\%) SO code snippets, while Gemini identified 120 (71\%). These results indicate that both models failed to detect at least 20\% of security issues. In the next step, we examined whether guided prompt could improve the ability of LLMs to detect security vulnerabilities more effectively. \\
We observed varying detection behaviors between the models, for instance, Listing~\ref{lst:SOexample} presents a Puppet manifest\footnote{In Puppet, snippets are commonly referred to as manifests.} sourced from SO\footnote{\url{https://stackoverflow.com/questions/23157739/mocking-methods-in-puppet-rspec-tests}}, which includes the security smell \textit{Use of HTTP without TLS}. 
ChatGPT
\footnote{\url{https://chatgpt.com/share/680ce01b-f3a4-8010-b786-5d07d6f090cd}} 
identified this smell in response to the prompt. In contrast, Gemini
\footnote{\url{https://g.co/gemini/share/c5eb8b988463}}
failed to detect the issue.


\begin{table*}[ht]
\centering
\caption{LLMs' security smell detection performance with simple and guided prompts on SO}
\label{tab:stack_overflow_recall}
\scalebox{0.65}{\begin{tabular}{
    l
    S[table-format=3.1] S[table-format=3.1]
    S[table-format=3.1] S[table-format=3.1]
    S[table-format=3.1] S[table-format=3.1]
    S[table-format=3.1] S[table-format=3.1]
}
\toprule
\multirow{2}{*}{\textbf{Security Smell}} &
\multicolumn{4}{c}{\textbf{General Prompt}} & \multicolumn{4}{c}{\textbf{Guided Prompt}} \\
\cmidrule(lr){2-5} \cmidrule(lr){6-9}
& \multicolumn{2}{c}{\textbf{ChatGPT}} & \multicolumn{2}{c}{\textbf{Gemini}}
& \multicolumn{2}{c}{\textbf{ChatGPT}} & \multicolumn{2}{c}{\textbf{Gemini}} \\
\cmidrule(lr){2-3} \cmidrule(lr){4-5} \cmidrule(lr){6-7} \cmidrule(lr){8-9}
& {Ansible (\%)} & {Puppet (\%)} & {Ansible (\%)} & {Puppet (\%)}
& {Ansible (\%)} & {Puppet (\%)} & {Ansible (\%)} & {Puppet (\%)} \\
\midrule
S1: Hard-coded secret  & 91.6 & 84.6 & 88.4 & 76.9 & 97.9 & 100.0 & 90.5 & 92.3 \\
S2: Empty password  & 22.2 & {NA} & 11.1 & {NA} & 100.0 & {NA} & 77.78 & {NA} \\
S3: Weak cryptography algorithm  & 50.0 & 80.0 & 50.0 & 20.0 & 100.0 & 100.0 & 100.0 & 80.0 \\
S4: Admin by default  & 62.5 & 66.7 & 62.5 & 33.3 & 43.8 & 0.0 & 56.3 & 0.0 \\
S5: HTTP without TLS  & 80.0 & 100.0 & 46.7 & 0.0 & 80.0 & 100.0 & 53.3 & 0.0 \\
S6: No integrity check  & 33.3 & {NA} & 100.0 & {NA} & 33.3 & {NA} & 0.0 & {NA} \\
S7: Invalid IP binding  & 0.0 & 100.0 & 0.0 & 25.0 & 0.0 & 100.0 & 100.0 & 50.0 \\
S8: Suspicious comments  & {NA} & {NA} & {NA} & {NA} & {NA} & {NA} & {NA} & {NA} \\
\midrule
\textbf{AVG} & \textbf{79.7} & \textbf{84.6} & \textbf{74.8} & \textbf{50.0} & \textbf{88.1} & \textbf{88.5} & \textbf{80.4} & \textbf{69.2} \\
\bottomrule
\end{tabular}}
\end{table*}

\emph{Guided Prompt Results.} 
To evaluate the impact of our guided prompt, we re-ran the security detection task using this revised prompt. ChatGPT's detection rate increased from 80\% to 88\%, while Gemini demonstrated a more substantial improvement, with its detection rate rising from 71\% to 78\%.

Table \ref{tab:stack_overflow_recall} shows the results of the general and the guided prompt for each security smell in Ansible and Puppet scripts.\vspace{2pt}

\emph{Fix suggestion.}
Beyond detection, we analyzed the extent to which ChatGPT and Gemini offer meaningful corrective suggestions or actionable guidance for identified security smells. ChatGPT provided explicit code fixes in 5\% of cases, slightly higher than Gemini’s 4\%. For broader recommendations, such as best practices or mitigation strategies, ChatGPT responded in 95\% of instances, while Gemini did so in 96\%. Both models consistently offered either a valid recommendation or a code correction for every detected issue.

\section{GitHub Study}
\label{GitStudy}

Motivated by the promising performance of LLMs in our SO study on detecting IaC security smells, especially with our guided prompt, we extend our investigation from SO code snippets to real projects. Using IaC scripts from GitHub, we evaluate LLMs’ capabilities for both detecting IaC misuses and generating secure IaC, aiming to derive deeper, actionable insights and advance our understanding of LLMs’ capabilities for securing IaC.


\emph{Study Scope.} Our study maintains the same scope and setting as the SO study.

\subsection{Data Collection}

We collected our GitHub data from state-of-the-art replication packages and direct searches on GitHub. 

\emph{From State-of-the-Art (SoTA) Literature.} 
We conducted a comprehensive review of the academic literature pertaining to security smells in IaC, specifically focusing on Ansible and Puppet scripts. Through this review, we identified and extracted 21,162 GitHub files that have been acknowledged in the literature for containing security vulnerabilities~\cite{rahman2021security, rahman2019systematic}.

\emph{Direct GitHub Search.}
To augment our dataset and enhance its comprehensiveness, we performed a systematic search directly on GitHub. This search was designed not only to expand our dataset but also to ensure the inclusion of previously unexamined data, which is crucial for evaluating LLMs. Our search criteria were based on the programming languages (Puppet and Ansible) and specific keywords, as detailed in Listing \ref{lst:Keyw}. This process initially yielded 1,369 GitHub code files, which were subsequently vetted through manual inspection, resulting in an additional 595 files identified as containing security smells.

In total, we collected 21,757 GitHub files.

\subsubsection{Sampling}
We selected a subset of the collected data for a manual inspection and study on LLMs. To ensure our findings are representative of the entire dataset, we included 430 GitHub files: 302 from SoTA, 128 from GitHub search. This sample exceeds the size required for 95\% confidence level with a 5\% margin of error.

\subsection{Experimental Design}
\label{sec:GitMethodology}

This section outlines a three-stage approach for assessing LLMs' ability in identifying and mitigating IaC security smells, using samples sourced from GitHub.

\subsubsection{Manual Checking and Categorization}
We conducted a lightweight, open-coding-like process during which two knowledgeable individuals independently and manually reviewed 430 GitHub IaC scripts. They categorized these scripts into nine distinct security smell categories, extracting unique patterns for each category. Next, they compared their result, discussing any disagreements to reach consensus. When differences of opinion persisted, a professor with relevant expertise was consulted. Ultimately, the results were finalized using a majority voting mechanism, achieving high inter-rater reliability (Cohen’s kappa = 0.89).

\subsubsection{security smell Detection}

To extend our findings from SO, we applied the same approach used for security smell detection in the SO study to scripts extracted from GitHub repositories. However, instead of developing a new prompt, we directly applied the guided prompt developed and validated in the SO study, ensuring methodological consistency and enabling cross-source comparison.

\subsubsection{Secure IaC Generation}

In many cases, developers lack pre-existing code and depend on LLMs to generate IaC scripts from scratch. This section assesses the capability of LLMs to produce secure IaC scripts when confronted with scenarios inspired by real-world examples extracted from GitHub.

\emph{Scenario Extraction (Direct Scenarios).} For this evaluation, two researchers collaboratively designed a scenario corresponding to each identified pattern from the categorization phase. Each scenario was crafted to prompt the LLM to generate code that might introduce security vulnerabilities, without explicitly stating that it is insecure. For instance, a scenario could request the use of SHA1 for password encryption, allowing an assessment of whether the LLM recognizes and addresses the weakness, or if it instead suggests a more secure alternative.

\subsection{Study Results}
\label{sec:result}
This section presents the results of our GitHub study. 
\subsubsection{Manual Categorization}

Through manual categorization of a sample dataset, we analyzed 430 IaC snippets, identifying security smells in 341 instances, while 89 snippets were found to be free of such issues. Table~\ref{tbl:SmellDistGitHub} shows the distribution of security smells in IaC scripts collected from GitHub. Notably, the “Missing Default in Case Statement” smell does not apply to Ansible, as it lacks switch or case constructs. Furthermore, this smell was not observed in either the Puppet data from Rahman et al.'s studies or within our own dataset.

By including both secure and insecure IaC scripts in our sample, we are able to evaluate key metrics, such as precision and F1 score, in the detection capabilities of LLMs, rather than being constrained to detection rate alone (as was the case in our SO study).

Additionally, we identified 89 unique security smell patterns among the 341 insecure GitHub IaC snippets. These patterns are used to generate 89 scenarios for assessing LLMs performance in secure IaC generation.

\begin{table}[ht]
\caption{Distribution of security smells on GitHub}
\centering
\scalebox{0.65}{\begin{tabular}{lcc}
\hline
\textbf{Security Smell} & \textbf{Ansible} & \textbf{Puppet} \\ \hline
\textbf{S1} & 109 & 45 \\
\textbf{S2} & 16 & 8 \\
\textbf{S3} & 2 & 1 \\
\textbf{S4} & 5 & 4 \\
\textbf{S5} & 27 & 17 \\
\textbf{S6} & 10 & 9 \\
\textbf{S7} & 8 & 5 \\
\textbf{S8} & 52 & 23 \\ \hline
\textbf{Total} & \textbf{229} & \textbf{112} \\ \hline
\end{tabular}}
\label{tbl:SmellDistGitHub}
\end{table}


\subsubsection{Detection}
In this section, we present the 
models' performance using simple and guided prompts.\vspace{1pt}

\emph{General Prompt Result.}
ChatGPT detected 145 (42\%) of the scripts, and Gemini identified 177 (51\%). These rates are notably lower than for SO posts, indicating both models missed many security issues in GitHub code.

Table \ref{tab:Gitsmell_detection} presents the results of  general and  guided prompts for each security smell in Ansible and Puppet scripts. ChatGPT achieved F1 scores of 53.1 for Ansible and 63.9 for Puppet, while Gemini achieved 57.5 and 78.7, respectively.

    
  

\begin{tcolorbox}[enhanced,
  attach boxed title to top left={yshift=-3mm,yshifttext=-1mm},
  colback=yellow!20,          
  colframe=yellow!20,         
]
When instructed to analyze the code from a security perspective, both ChatGPT and Gemini failed to detect more than half of the security smells.
\end{tcolorbox}

\begin{table*}[ht]
\centering
\caption{LLMs' security smell detection performance with simple prompt on GitHub}
\label{tab:Gitsmell_detection}
\scalebox{0.63}{
\begin{tabular}{
    l
    S[table-format=2.1] S[table-format=3.1] S[table-format=3.1]
    S[table-format=2.1] S[table-format=3.1] S[table-format=3.1]
    S[table-format=2.1] S[table-format=3.1] S[table-format=3.1]
    S[table-format=2.1] S[table-format=3.1] S[table-format=3.1]
}
\toprule
\multirow{2}{*}{\textbf{\begin{tabular}[c]{@{}c@{}}Security \\   Smell\end{tabular}}} &
\multicolumn{6}{c}{\textbf{ChatGPT}} &
\multicolumn{6}{c}{\textbf{Gemini}} \\
\cmidrule(lr){2-7} \cmidrule(lr){8-13}
& \multicolumn{3}{c}{\textbf{Ansible}} & \multicolumn{3}{c}{\textbf{Puppet}} 
& \multicolumn{3}{c}{\textbf{Ansible}} & \multicolumn{3}{c}{\textbf{Puppet}} \\
\cmidrule(lr){2-4} \cmidrule(lr){5-7} \cmidrule(lr){8-10} \cmidrule(lr){11-13}
& {\textbf{Recall (\%)}} & {\textbf{Precision (\%)}} & {\textbf{F1 (\%)}} 
& {\textbf{Recall (\%)}} & {\textbf{Precision (\%)}} & {\textbf{F1 (\%)}}
& {\textbf{Recall (\%)}} & {\textbf{Precision (\%)}} & {\textbf{F1 (\%)}}
& {\textbf{Recall (\%)}} & {\textbf{Precision (\%)}} & {\textbf{F1 (\%)}} \\
\midrule
S1  & 37.6 & 85.4 & 52.2 & 48.9 & 95.7 & 64.8 & 25.0 & 88.2 & 67.6 & 86.7 & 100.0 & 92.8 \\
S2  & 68.8 & 91.7 & 78.6 & 50.0 & 100.0 & 66.7 & 25.0 & 100.0 & 40.0 & 37.5 & 100.0 & 54.5 \\
S3  & 50.0 & 100.0 & 66.7 & 100.0 & 100.0 & 100.0 & 66.7 & 100.0 & 80.0 & 100.0 & 100.0 & 100.0 \\
S4  & 40.0 & 25.0 & 30.8 & 50.0 & 100.0 & 66.7 & 100.0 & 41.7 & 58.8 & 75.0 & 100.0 & 85.7 \\
S5  & 77.8 & 91.3 & 84.0 & 88.2 & 100.0 & 93.7 & 59.3 & 94.1 & 72.5 & 82.4 & 93.3 & 87.5 \\
S6  & 50.0 & 100.0 & 66.7 & 44.4 & 100.0 & 61.5 & 60.0 & 85.7 & 70.6 & 88.9 & 100.0 & 94.1 \\
S7  & 87.5 & 100.0 & 93.3 & 60.0 & 100.0 & 75.0 & 62.5 & 100.0 & 76.9 & 20.0 & 100.0 & 33.3 \\
S8  & 11.5 & 100.0 & 20.6 & 4.3  & 100.0 & 8.2  & 5.8  & 100.0 & 11.0 & 34.8 & 100.0 & 51.6 \\
\midrule
\textbf{AVG} & \textbf{41.0} & \textbf{85.5} & \textbf{53.1} 
             & \textbf{46.4} & \textbf{98.1} & \textbf{63.9} 
             & \textbf{43.7} & \textbf{85.5} & \textbf{57.5} 
             & \textbf{68.8} & \textbf{98.7} & \textbf{78.7} \\
\bottomrule
\end{tabular}}
\end{table*}


\begin{table*}[ht]
\centering
\caption{LLMs' security smell detection performance with guided prompt on GitHub}
\label{tab:prompt_engineering_results_with_f1}
\scalebox{0.63}{
\begin{tabular}{
    l
    S[table-format=2.1] S[table-format=3.1] S[table-format=3.1]
    S[table-format=2.1] S[table-format=3.1] S[table-format=3.1]
    S[table-format=2.1] S[table-format=3.1] S[table-format=3.1]
    S[table-format=2.1] S[table-format=3.1] S[table-format=3.1]
}
\toprule
\multirow{2}{*}{\textbf{\begin{tabular}[c]{@{}c@{}}Security \\   Smell\end{tabular}}} &
\multicolumn{6}{c}{\textbf{ChatGPT}} & \multicolumn{6}{c}{\textbf{Gemini}} \\
\cmidrule(lr){2-7} \cmidrule(lr){8-13}
& \multicolumn{3}{c}{\textbf{Ansible}} & \multicolumn{3}{c}{\textbf{Puppet}} 
& \multicolumn{3}{c}{\textbf{Ansible}} & \multicolumn{3}{c}{\textbf{Puppet}} \\
\cmidrule(lr){2-4} \cmidrule(lr){5-7} \cmidrule(lr){8-10} \cmidrule(lr){11-13}
& {\textbf{Recall (\%)}} & {\textbf{Precision (\%)}} & {\textbf{F1 (\%)}} 
& {\textbf{Recall (\%)}} & {\textbf{Precision (\%)}} & {\textbf{F1 (\%)}}
& {\textbf{Recall (\%)}} & {\textbf{Precision (\%)}} & {\textbf{F1 (\%)}}
& {\textbf{Recall (\%)}} & {\textbf{Precision (\%)}} & {\textbf{F1 (\%)}} \\
\midrule
S1 & 89.9 & 84.5 & 87.1 & 100.0 & 95.7 & 97.8 & 82.6 & 82.6 & 82.6 & 91.1 & 100.0 & 95.3 \\
S2 & 93.8 & 93.8 & 93.8 & 75.0  & 100.0 & 85.7 & 50.0 & 88.9 & 64.0 & 50.0 & 100.0 & 66.7 \\
S3 & 50.0 & 100.0 & 66.7 & 100.0 & 100.0 & 100.0 & 0.0 & 0.0 & 0.0 & 100.0 & 100.0 & 100.0 \\
S4 & 20.0 & 6.7 & 10.2 & 50.0 & 66.7 & 57.1 & 20.0 & 5.6 & 8.9 & 75.0 & 75.0 & 75.0 \\
S5 & 88.9 & 92.3 & 90.5 & 100.0 & 94.4 & 97.1 & 0.0 & 0.0 & 0.0 & 64.7 & 84.6 & 73.1 \\
S6 & 30.0 & 50.0 & 37.5 & 88.9 & 100.0 & 94.1 & 40.0 & 50.0 & 44.4 & 66.7 & 75.0 & 70.6 \\
S7 & 100.0 & 88.9 & 94.1 & 100.0 & 100.0 & 100.0 & 50.0 & 100.0 & 66.7 & 100.0 & 100.0 & 100.0 \\
S8 & 98.1 & 100.0 & 99.0 & 91.3 & 100.0 & 95.5 & 53.8 & 100.0 & 69.9 & 87.0 & 100.0 & 93.1 \\
\midrule
\textbf{AVG} & \textbf{87.8} & \textbf{83.8} & \textbf{85.5}
            & \textbf{93.8} & \textbf{96.3} & \textbf{90.6}
            & \textbf{59.0} & \textbf{75.4} & \textbf{47.5}
            & \textbf{81.3} & \textbf{94.8} & \textbf{84.2} \\
\bottomrule
\end{tabular}}
\end{table*}


\begin{lstlisting}[caption={Example of suspecious comments on GitHub}, label={lst:Gitexample1}, breaklines=true]
xqwatcher_repo_name: xqueue-watcher.git
 #TODO: change this to /edx/etc after pulling xqwatcher.json out
xqwatcher_conf_dir: "{{ xqwatcher_app_dir }}"
#TODO: remove after refactoring out all the git stuff
xqwatcher_course_git_ssh_opts: "-o UserKnownHostsFile=/dev/null -o StrictHostKeyChecking=no -i {{ xqwatcher_app_dir }}/.ssh/{{ xqwatcher_service_name }}-courses"
xqwatcher_requirements_file:"{{ xqwatcher_code_dir }}/requirements.txt"
xqwatcher_log_dir: "{{ COMMON_LOG_DIR }}/{{ xqwatcher_service_name }"
xqwatcher_module:"xqueue_watcher"

\end{lstlisting}



For example, Listing~\ref{lst:Gitexample1} shows an Ansible playbook\footnote{In Ansible, code snippets are commonly referred to as playbooks.} retrieved from GitHub\footnote{\url{https://github.com/SLIC-Ansible/blob/master/repo-github/pedrorib40istx/playbooks/roles/xqwatcher/defaults/main.yml}}, which contains the security smell \textit{Suspicious Comment (TODO)}.
 This snippet, associated with the deployment of the \texttt{xqwatcher} server, appears functionally correct but includes insecure configurations: line 5 disables SSH host verification using \texttt{StrictHostKeyChecking=no} and \texttt{UserKnownHostsFile=/dev/null}, which may allow man-in-the-middle (MITM) attacks. Moreover, a \texttt{\#TODO} comment in line 4 suggests incomplete or temporary implementation, potentially aiding attackers by revealing 
 weaknesses.

In this case, both ChatGPT\footnote{\url{https://rb.gy/3ywgas}} and Gemini\footnote{\url{https://g.co/gemini/share/252f16ab47dc}} failed to identify the security smell. Furthermore, both models exhibited generally lower detection rates for suspicious comments across the dataset, underscoring the complexity of reliably identifying this category of smells.

\emph{Guided Prompt result.}
Table~\ref{tab:prompt_engineering_results_with_f1} presents the detailed results of the detection of security smells after applying the guided prompt. The findings demonstrate a significant improvement in detection performance for both models, with the impact being especially pronounced for GPT. GPT’s F1 score increased substantially from 58\% to 89\%, while Gemini’s F1 score showed 
modest improvement, rising from 66\% to 74\%. 
\begin{tcolorbox}[enhanced,
  attach boxed title to top left={yshift=-3mm,yshifttext=-1mm},
  colback=yellow!20,          
  colframe=yellow!20,         
]

In summary, guided prompt significantly enhanced security smell detection in both models, with GPT showing a more notable improvement. However, many end users may not know how to construct effective prompts. This highlights the responsibility of researchers and developers to raise awareness and share effective prompts for common tasks.
\end{tcolorbox}

\emph{Fix suggestion.}
ChatGPT offered explicit code corrections in 15\% of cases, compared to only 5\% for Gemini. Regarding general recommendations
, ChatGPT responded in 83\% of cases, while Gemini gave recommendations in 93\% of instances. In 2\% of cases, both models failed to provide any response beyond detection.

\subsubsection{Generation}
Since developers frequently utilize LLMs as assistant tools to generate code for specific use cases, we examined how LLMs respond when prompted to generate code for scenarios that are likely to introduce security smells. Specifically, we investigated the extent to which LLMs issue warnings or guidance in such situations. For this purpose, we analyzed 89 synthetic, deliberately insecure scenarios, designed to reflect a wide range of potential vulnerabilities and covering the entirety of the 341 GitHub code samples.
    
  


\begin{figure}
  \centering
  \begin{minipage}{0.24\textwidth}
    \centering
    \includegraphics[width=\textwidth]{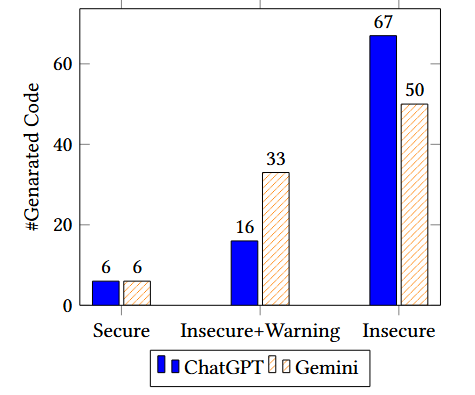}
    (a) 
    \label{fig:scenarios}
  \end{minipage}
  \hfill
  \begin{minipage}{0.24\textwidth}
    \centering
    \includegraphics[width=\textwidth]{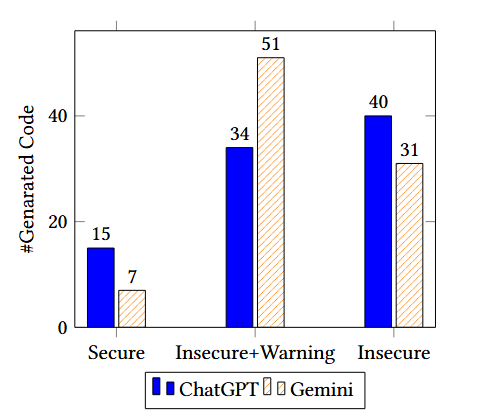}
    (b)
    \label{fig:scenariosSec}
  \end{minipage}
  \caption{LLM's performance for code generation. In (a) we did not ask for ``secure'' code, whereas in (b) we asked explicitly.}
  \label{fig:combined}
\end{figure}

As illustrated in Figure~\ref{fig:combined}(a), both ChatGPT and Gemini frequently generate insecure code when prompted with scenarios that are prone to security smells, often without issuing any warnings. Specifically, 75\% of the code generated by ChatGPT and 56\% by Gemini contained security smells and lacked any warning message. In 17\% of the cases, ChatGPT generated insecure code but included a warning indicating that the code was not secure, while Gemini did so in 37\% of the cases. Alarmingly, both models produced fully secure code in only 7\% of the scenarios, highlighting a substantial gap in security-aware code generation.
Figure~\ref{fig:combined}(b) presents the results of prompting both ChatGPT and Gemini to generate secure code by explicitly including the keyword \textit{“secure”} in the request. Despite this precaution, 44\% of the code generated by ChatGPT and 34\% by Gemini still contained security smells without any accompanying warnings. This finding suggests that even when developers explicitly express concern for security, the generated IaC may remain vulnerable, underscoring the limitations of current LLMs in reliably enforcing security constraints during code generation.

    
  


\begin{tcolorbox}[enhanced,
  attach boxed title to top left={yshift=-3mm,yshifttext=-1mm},
  colback=yellow!20,          
  colframe=yellow!20,         
]
Both models generate secure IaC for only 7\% of synthetic scenarios, and even when explicitly requested to produce “secure” code, at most 19\% of the outputs were secure, suggesting that LLMs may give developers a false sense of security.
\end{tcolorbox}

For example, consider the following scenario, emphasizing secure code practice: \textit{``I have an Ansible playbook for automating server management and application deployment, as shown below:}

\begin{lstlisting}[label={lst:22},frame=none,numbers=none]
dest: /etc/yum.repos.d/atomic7-testing.repo
content: |
[atomic7-testing]
name=CentOS7 Atomic Testing
baseurl=http://cbs.centos.org/repos/atomic7-testing/x86_64/os/
gpgcheck=0
enabled=1
exclude=librepo
\end{lstlisting}
 \textit{Please generate a new \textbf{secure code} to ensure that Docker and libvirt daemon services are started and that the system fetches fresh metadata from repositories.''}
 
Upon reviewing ChatGPT's 
response
, several security concerns are evident. Specifically, we observe \emph{S6 (No integrity check)} smell due to the disabled GPG and a potential \emph{S5 (HTTP without TLS)}. Furthermore, the unconditional use of root privileges (\texttt{become: yes}) across all tasks can be seen as an instance of \emph{S4 (Admin by default)}, and it did not raise any warning against them.


    
  

\section{Discussion}
\label{sec:Discussion}

\emph{\textbf{- Insights into Analyzing IaC snippets with LLMs.}}\vspace{1pt}

\emph{Puppet vs. Ansible.}
Both models generally exhibit superior performance on \emph{Puppet} scripts compared to \emph{Ansible} scripts. This trend is particularly evident in the precision and recall metrics for \emph{S1 (Hard-coded secret)}, \emph{S3 (Weak cryptography algorithm )}, and \emph{S5 (HTTP without TLS)}. However, an exception is observed with \emph{S7 (Invalid IP binding)}, where both models demonstrate enhanced detection capabilities for Ansible code.
This discrepancy may be attributed to the structural differences between the languages. Puppet's more declarative and rigid format may aid LLMs in pattern recognition, whereas Ansible’s YAML-based structure allows for greater flexibility and variability, complicating the detection of code smells.

\emph{Stack Overflow vs. GitHub.}
When comparing datasets, both models perform better on SO  compared to GitHub, especially when using general prompts. This outcome is likely due to the inherent nature of GitHub IaC scripts, which frequently include variables and configuration metadata, in contrast to the often partial and simplified examples found on SO.

\emph{Variability in Security Smell Detection.}
No consistent trend is observed in the detection of each security smell, as performance varies across them. Notably, the detection of \emph{S8 (Suspicious comments)} is particularly low, whereas \emph{S5 (HTTP without TLS)} shows relatively good performance.

\emph{The Role of a guided prompt.}
While both models benefit from guided prompt, the improvement is considerably more pronounced for ChatGPT.
 Guided prompt significantly enhances the detection of \emph{S8 (Suspicious comments)} and \emph{S1 (Hard-coded secrets)}, highlighting its effectiveness in textual pattern recognition.
 Conversely, it negatively impacts the detection of \emph{S4 (Admin by default)} and \emph{S6 (No integrity check)}.\vspace{2pt}

\emph{\textbf{- IaC Platforms.}}
In our manual checking process, we systematically identified and documented the names of relevant IaC platforms based on their configured scripts. These platforms were classified into eight distinct categories: 
\emph{P1: Database/Storage Solution,
 P2: Cloud and Infrastructure Platform,
 P3: Monitoring, Logging, and Analytics,
 P4: Development and CI/CD,
 P5: Security and Compliance Solutions,
 P6: Messaging and Queue Platform,
 P7: Educational and Training Platforms, and
 P8: Other/Utility Platforms. }\vspace{2pt}
 
\begin{figure}
  \centering
\includegraphics[width=.31\textwidth]{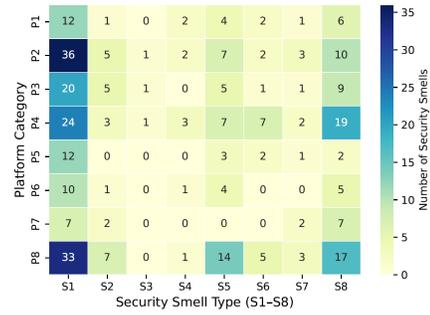}
\caption{Distribution of security smells in each platform category}
\label{fig: Heat}
\end{figure}

 Figure \ref{fig: Heat} shows a heat map of the distributions of security smells across categories, highlighting their prevalence on each type of platform.
 
Our analysis reveals that both ChatGPT and Gemini excelled in the \textit{Database} category (P1) using general prompt, with detection rates of 57.14\% and 60.71\% respectively, highlighting their strong performance in structured data contexts.

In contrast, initial detection rates within the \textit{Security and Compliance} (P5) and \textit{Messaging and Queue} (P6) categories were lower for ChatGPT, at 25\% and 27\%. Notably, after applying prompt engineering, detection rates in these categories increased significantly to 85\% and 90\%, demonstrating the effectiveness of guided prompt.

\vspace{2pt}

\emph{\textbf{- Code Similarity.}}
 To explore potential bias in LLMs and their resemblance to training sources, a resemblance that might inadvertently replicate insecure code, we reviewed SO posts with accepted answers containing code snippets
. We then assessed how closely the responses from ChatGPT and Gemini aligned with the accepted solutions.

Similarity was assessed using (i) a manual 1–4 scale (1 = not similar, 2 = somewhat similar, 3 = very similar, 4 = exact match) and (ii) an automated text-based method (TF–IDF with cosine similarity) applied to raw code snippets, without analyzing structural properties of the code. Both methods produced consistent patterns. Manual ratings (means): Gemini–SO = 1.90, ChatGPT–SO = 2.09, Gemini–ChatGPT = 2.90. Automated scores: Gemini–SO = 41.39\%, ChatGPT–SO = 41.24\%, Gemini–ChatGPT = 72.54\%.

These findings suggest that, for each SO question, code snippets generated by the LLMs resemble each other more than they resemble the accepted SO answer. Nonetheless, this limited study provides only preliminary insight; further research is needed to examine broader datasets, structural code similarity, and implications for code quality and security.

\section{Threats to Validity}
\label{sec:Threats}

\emph{Internal Validity.}
There are several threats to internal validity that may potentially affect the soundness of our study.

 \emph{1. Human Bias in Categorization:} Manual data checking can introduce bias. To mitigate this, each entry was reviewed by at least two individuals, and disagreements were resolved through majority voting.
 
 \emph{2. Potential Data Leakage:} Using datasets from SoTA may overlap with the training data of LLMs. We mitigated this by including newly gathered data.
 
 \emph{3. Response Variability:} LLM responses can vary with identical prompts. Our dataset includes multiple code snippets per security smell to ensure a reliable assessment of LLM’s detection ability.\vspace{2pt}
 
\emph{External Validity.}
Threats to external validity arise from concerns about the generalizability of our study’s results.\vspace{2pt}

\emph{1. Specificity to IaC Tools:} The study's findings are specific to detecting IaC security smells in Puppet and Ansible scripts. This may limit generalizability to other IaC tools.

 \emph{2. Model Limitations:} Our evaluation focuses on GPT-4o and Gemini 2.0 flash models, which may not represent the capabilities of other language models.
    

\section{Related Work}
\label{sec:RelatedWork}
Prior work relevant to our study falls into two areas: (1) security smells in IaC and (2) recent LLM-based security analysis and secure code generation.

\emph{IaC Security Smells.}
The exploration of security in In IaC has evolved significantly over the years, 
In 2016, Sharma et al.~\cite{sharma2016does} provided an influential catalog detailing 13 implementation and 11 design configuration smells specific to IaC, analyzing 8.9 million lines of code from 4,621 Puppet repositories. They found that design smells co-occur 9\% more often than implementation smells. Building on this, Schwarz et al.~\cite{schwarz2018code} in 2018 identified 17 technology-agnostic code smells applicable to Chef and Puppet, demonstrating their wide prevalence across diverse IaC tools.

The foundational work by Rahman et al.~\cite{rahman2019seven,rahman2021security,rahman2019systematic,RahmanPup}, spanning from 2019 to 2023, highlighted the widespread occurrence of security smells across Puppet, Ansible, and Chef environments. Their studies introduced tools like SLIC, SLAC, and TaintPup.
They revealed that 75\% of IaC files contained at least one smell, with some persisting for up to 98 months.




\emph{LLMs for Security.}
 Research has explored both the capabilities and limitations of LLMs in enhancing security~\cite{ srivatsa2024survey,asare2024user, perry2023users, firouzi2024time, Kavian2024-lv, FIROUZIGenAI, Soltaniani2026DataLeakSecret,Firouzi2026Securecode}. 

LLMs have been widely studied for vulnerability detection, but the results are inconsistent. 
Steenhoek et al.~\cite{steenhoek2024comprehensive} reported a 57\% error rate in LLM-generated vulnerability explanations, 
while Lucas et al.~\cite{lucas2024evaluating} found GPT-4 identified software test smells with 70\% accuracy.

Concerns about the generation of insecure code by LLMs have also been raised. Srivatsa et al.~\cite{srivatsa2024survey} noted that LLMs sometimes produce insecure yet seemingly correct code due to constraints in their training datasets. 

Studies~\cite{ChatGPTMisuseDetector2024,Prompting} also show that prompt engineering can enhance vulnerability detection and secure code generation. 
While interest in security and LLMs is growing, little research examines their effectiveness in detecting security smells in IaC. Our work addresses this gap by evaluating how well LLMs detect security smells and generate secure IaC scripts.

\section{Conclusion}
\label{sec:Conclusion}

We investigated the support of Large Language Models (LLMs) for secure Infrastructure as Code (IaC).
When we only instructed models to detect security risks, more than half of the security smells in GitHub IaC scripts remained undetected. 
With a guided prompt, detection accuracy and F1 scores increased substantially. 
When we instructed models to generate IaC scripts for vulnerable synthetic scenarios, they produced scripts with security smells and mostly with no warnings.
We observed a non-promising performance even when we asked LLMs to generate secure solutions.
These findings highlight the need for further research to leverage LLMs' potential for secure IaC development.



\bibliographystyle{IEEEtran}
\bibliography{sample-base}
\end{document}